\documentclass[prb, twocolumn, showpacs]{revtex4}
\usepackage{graphicx}

\begin{document}
\title{Search for Cooper-pair Fluctuations in Severely Underdoped Y$_{1-y}$Ca$_y$Ba$_2$Cu$_3$O$_{6+x}$ Films ($y=0$ and $0.2$)}
\author{Yuri L. Zuev}
\thanks{Present address: Materials Science and Technology Division, Oak Ridge National Laboratory, Oak Ridge, TN}
\author{Iulian Hetel and Thomas R. Lemberger}
\affiliation{Department of Physics, The Ohio State University\\ 191 W. Woodruff Ave., Columbus, OH, 43210}

\begin{abstract}
The preformed-pairs theory of pseudogap physics in high-$T_C$ superconductors predicts a nonanalytic $T$-dependence for the $ab$-plane superfluid fraction, $\rho_S$, at low temperatures in underdoped cuprates. We report high-precision measurements of $\rho_S(T)$ on severely underdoped YBa$_2$Cu$_3$O$_{6+x}$ and Y$_{0.8}$Ca$_{0.2}$Ba$_2$Cu$_3$O$_{6+x}$ films. At low $T$, $\rho_S$ looks more like $1 - T^2$ than $1 - T^{3/2}$, in disagreement with theory.
\end{abstract}

\maketitle

Numerous theoretical models have been advanced to explain the pseudogap in underdoped high-$T_C$ superconductors. One class of models ({\em boson-fermion models})~\cite{Geshkenbein, Micnas, Micnas1} involves the idea that Cooper pairs form at the pseudogap temperature, $T^*$, but due to internal fluctuations, these pairs do not form a condensate until the observed transition temperature, $T_C$, where superfluid appears. Of the many results of the model, (\cite{Stajic} and references therein) to us the most dramatic and accessible is its prediction of a nonanalytic $T$-dependence of the superfluid fraction - a quantity that we can measure with high precision. In this model, the pairing field $\Delta_{\mathrm{pg}}(T)$ binds the carriers into pairs below a pseudogap temperature, $T^*$, of the order of several hundred K. The $T$ dependence of $\Delta_{\mathrm{pg}}$ is assumed to be BCS-like. Below $T_C$, an energy scale $\Delta_{\mathrm{sc}}$ responsible for pair condensation appears. The total gap, $\Delta (T)$, for Fermionic quasiparticles is: $\Delta^2=\Delta_{\mathrm{pg}}^2+\Delta_{\mathrm{sc}}^2$. The $T$ dependence of the superfluid density is determined by condensation of preformed pairs. The underlying BCS superfluid fraction, $\rho_s^{\mathrm{BCS}}(T)$, which vanishes at $T^*$, is renormalized by pair condensation such that the measured superfluid fraction is: $\rho_s(T)=[\Delta_{\mathrm{sc}}^2(T)/\Delta^2(T)]\rho_s^{\mathrm{BCS}}(T)$. 

Detailed calculations~\cite{Kosztin, Chen1} predict that strong Cooper pair fluctuations in underdoped materials dominate the physics and give $\Delta_{\mathrm{sc}}(T)$, and therefore $\rho_s(T)$, a nonanalytic $T$-dependence: $\Delta_{\mathrm{sc}}^2(T) \propto 1-a(T/T_C)^{3/2}$, where $a$ is near unity. While this is technically a low-$T$ result, it is expected to hold over a relatively wide temperature range because $T_C$ is the only low-energy scale in the problem. Because $T^*$ is much larger than $T_C$, $\Delta_{\mathrm{pg}}$ is essentially constant at low $T$, $T \ll T^*$. And, in moderately disordered $d$-wave superconductors, $\rho_s^{\mathrm{BCS}}(T)$, is quadratic at low $T$~\cite{Hirschfeld}. The end result is: $\rho_s(T) \approx 1-a(T/T_C)^{3/2}- b(T/T^*)^2$, where $b$ is a constant near unity~\cite{Chen2}. This result is experimentally accessible for two reasons. First, there is the separation of scales: for the severe underdoping explored here, $T^*$ is more than ten times larger than $T_C$, in which case the quadratic term is negligible. Second, the predicted nonanalytic $T$-dependence is not affected by microscopic disorder, which is present in our films. Numerical results~\cite{Chen2} on disordered d-wave superconductors find that the limiting behavior, $\rho_s(T) \approx 1 - a(T/T_C)^{3/2}$, holds up to about $T_C/2$.

The theory under discussion is physically appealing, but controversial. Rather than discuss its merits, we proceed to test its striking prediction for $\rho_S(T)$ in cuprates. We note that Carrington et al.~\cite{Carrington} claim $T^{3/2}$ behavior in organic superconductors. In this brief report, we present the $T$-dependence of $\rho_S(T)$ for several heavily underdoped YBCO films. The discussion is couched in terms of the $ab$-plane penetration depth, $\lambda$, where $\rho_s(T) = \lambda^{-2}(T)/ \lambda^{-2}(0)$. None of the films exhibits a clear $T^{3/2}$ dependence over a significant temperature range. The data are better fit by $T^2$. 

We determine the $ab$-plane superfluid density of thin films by a two-coil {\em rf} technique, described in detail in~\cite{Turneaure1, Turneaure2}. The measured quantity is the sheet conductivity of the film, $\sigma d = \sigma_1 d-i\sigma_2 d$, where $d$ is the film thickness and $\sigma$ is the usual conductivity. The $ab$-plane penetration depth, $\lambda$, is obtained as: $\lambda^{-2}=\mu_0\omega\sigma_2$, where $\mu_0$ is magnetic permeability of vacuum and $\omega/ 2\pi$ is the measurement frequency, 100kHz in this work. $\sigma_1$ is large enough to be measurable only near the transition, where, in principle, it diverges. The width of this peak provides an upper limit on the inhomogeneity of $T_C$.

In this study we used three films of pure YBCO with $T_C = $ 36, 20 and 10 K, and $\lambda_{ab}(0)=$0.46, 0.75 and 2.43 $\mu$m, respectively. These are films A, B, and C in Table I. They were produced by pulsed laser deposition (PLD) on (001) polished SrTiO$_3$ substrates with subsequent annealing. Films were sandwiched between a buffer and a cap layer of insulating PrBCO, whose purpose was to improve homogeneity of YBCO by removing interfaces with substrate and air (protecting samples from air humidity). In addition, the PrBCO cap helps move the region of high oxygen concentration gradient (near the free surface) away from YBCO, thus improving oxygen homogeneity. The doping homogeneity is paramount here because at the doping levels of this work the $T_C$ is a strong function of doping, so that small variations of oxygen concentration cause wide transition and potentially much of the superfluid density curve $\lambda^{-2}(T)$ is inside the transition region. The above procedure allows us to achieve narrow enough $\sigma_1$ peaks so that no more than about a quarter of $\rho_s(T)$ curve is inside the peak. A fourth film (D) was 140 nm thick film of Y$_{0.8}$Ca$_{0.2}$BCO produced by PLD on (001) STO.

\begin{table}[b]
\begin{tabular}{|c|c|c|c|c|c|}\hline
Sample&Thickness, $\mu$m & $T_C$,K&$\lambda(0)$, $\mu$m&$X^2_{T^{3/2}}$&$X^2_{T^2}$\\ \hline
A & 0.048 & 36 & 0.45 & $3.0\cdot 10^{-5}$ &  $4.6\cdot 10^{-6}$   \\
B & 0.048 & 20 & 0.76 & $2.0\cdot 10^{-5}$ &  $1.2\cdot 10^{-6}$   \\
C & 0.025 & 10 & 2.4  & $8.3\cdot 10^{-5}$ &  $3.6\cdot 10^{-5}$   \\
D & 0.14  & 28 & 0.94 & $5.2\cdot 10^{-5}$ &  $5.8\cdot 10^{-6}$   \\ \hline
\end{tabular}
\label{tab:param}
\caption{Parameters for the films presented in this report. $X^2$ comes from fits to the first 20\% drop in $\lambda^{-2}(T)$.}
\end{table}

\begin{figure}[t]
\centering
\includegraphics[width=\columnwidth]{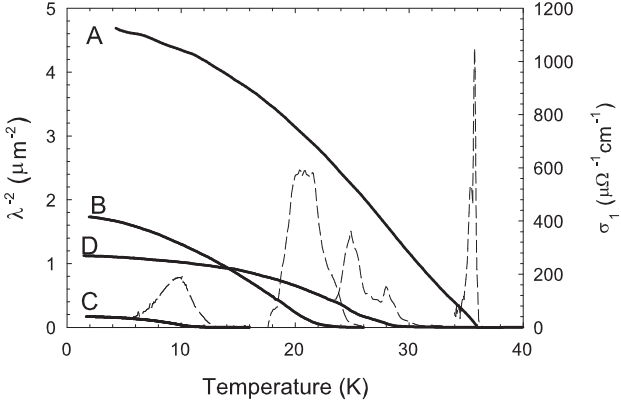}
\caption{Superfluid density $\rho_s\propto\lambda^{-2}$ (solid curves) and $\sigma_1$ (dashed curves) {\em vs.} $T$ for four films used in the present study.}
\label{fig:all}
\end{figure}

\begin{figure}[t]
\centering
\includegraphics[width=\columnwidth]{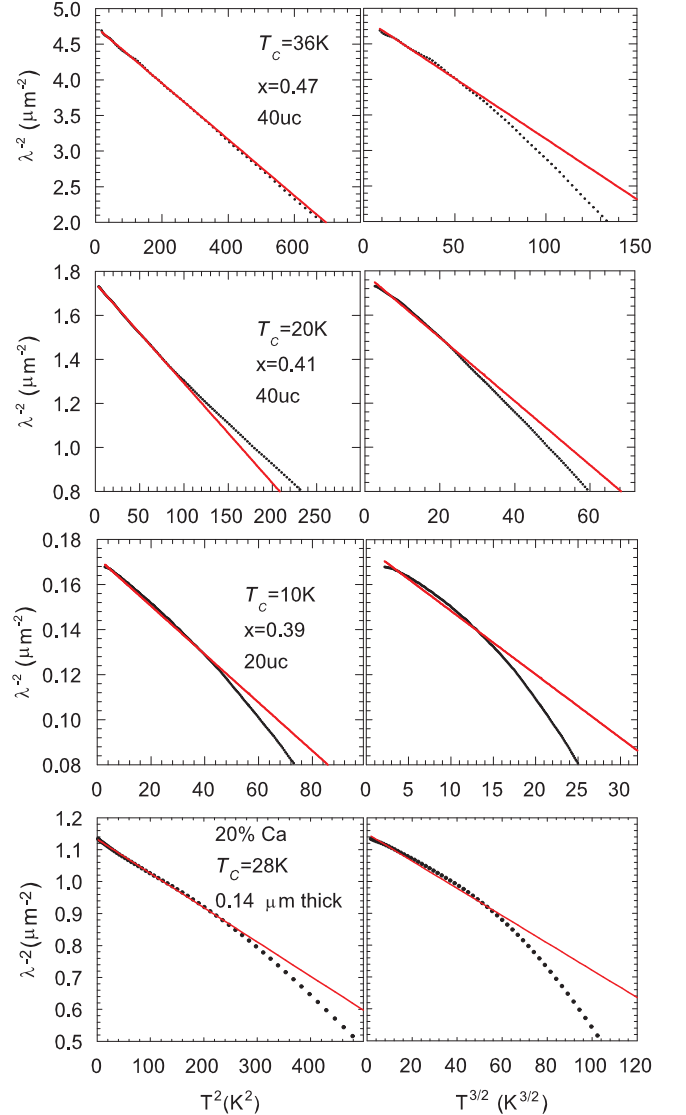}
\caption{(Color online) Low-$T$ part of the superfluid density for films A, B, C and D (top to bottom). In each pair of graphs, left one shows data against $T^2$, while the right one is against $T^{3/2}$. The red lines are linear fits to the first 20\% decrease of the superfluid density. In each case, quadratic $T$ dependence provides better fit over larger temperature range.}
\label{fig:ABCD}
\end{figure}


The main systematic uncertainty in our measurement is the uncertainty in the film thickness, $d$, which affects the absolute value of $\lambda^{-2}$ at $T=0$, but not its $T$ dependence. The 36 and 20 K films were nominally 40 unit cells thick ($d=0.048\mu$m), while the 10K film was 20 unit cells thick ($d=0.025\mu$m). We know the film thickness to 10\%, therefore we know $\lambda^{-2}(0)$ to within 10\% , and $\lambda (0)$ to 5\%. This uncertainty does not affect the conclusions of the present work. 

Figure~\ref{fig:all} shows the superfluid fraction, $\rho_s\propto\lambda^{-2}$, along with peaks in $\sigma_1$, as functions of $T$. The narrowest transition is that of the 36 K film. This film qualifies as "severely" underdoped because the pseudogap temperature is greater than 300 K~\cite{Timusk}.

In figure~\ref{fig:ABCD} we compare $T^2$ power law vs. $T^{3/2}$ law for our films. The black dotted curves are experimental data, while red lines are the power law fits to the first 20\% drop of the superfluid density. Left panels are the data plotted vs. $T^2$, while right panels are the data plotted against $T^{3/2}$. The two power laws are close to each other and are hard to distinguish. Yet, as far as we can tell, the quadratic fit is better over larger temperature interval. This is most evident in 36K film (top panel, fig.\ref{fig:ABCD}), where the quadratic fit is almost perfect over as much as 50\% of the superfluid density drop and 70\% of the temperature interval. 
 
To be more specific, we can describe the fit quality by calculating a fractional fit error:
\begin{equation}
X^2=\frac{1}{N}\sum\frac{(\rho_{s,\mathrm{fit}}-\rho_{s,\mathrm{data}})^2}{\rho_{s,\mathrm{fit}}^2}
\label{eq:chi2}
\end{equation}
over first 20\% of the drop in superfluid density $\rho_s$, where $N$ is the number of data points in that interval. Values of $X^2$ are given in the table for both power laws. One can see that $X^2$ is about an order of magnitude smaller for $T^2$ fit (good to about 0.1\%), than for $T^{3/2}$ fit, which is good only to about 1\% over same temperature interval. The quadratic fit quality at lowest temperatures for sample A is actually slightly worse than that for sample B, even though at higher temperatures sample A fits the $T^2$ law better.  
 
However, $T^{3/2}$ gives a reasonable agreement with data over some temperature range (10-15K). In principle, the fit can be improved somewhat by including $T^2$ term. However, as mentioned above, the quadratic term should be 100 times smaller than the $T^{3/2}$ term because $T^*$ is about 10 times larger than $T_C$.

One can be concerned about possible modification of the temperature dependence by spread in $T_C$ across a given sample. It is easy to show that such spread would cause only a change in a prefactor, but no new power law behavior, as long as the temperature is below about a half of the lowest $T_C$, which is the case here.

While the 36K film provides the cleanest comparison  with "pair fluctuation" theory because of its very sharp transition, the same qualitative conclusion is supported by data on 20 and 10 K films, as well as Ca doped film (three lower panel, fig.~\ref{fig:ABCD}). The transitions in these films are not narrow enough to draw more quantitative conclusions from those data.

It should be noted that if our films were cleaner, i.e. if they displayed $T$-linear dependence of $\rho_s$, it would be more difficult to reject the pair-fluctuations picture. In very clean crystals of D. Broun et al.~\cite{Broun} where $T$-linear dependence clearly yields to $T^2$ at lower temperatures, these two power laws together produce a good fit to $T^{3/2}$ behaviour over as much as 40\% drop in superfluid density. We believe that this is accidental and is not caused by preformed pairs condensation.

In conclusion, we made high precision measurements of superfluid density in heavily underdoped YBCO to test a novel prediction of a particular model of the pseudogap. We did not find persuasive evidence for the predicted $1 - a(T/T_C)^{3/2}$ behavior in superfluid fraction at low $T$. In every case, a quadratic $T$ dependence fitted the data better. It is possible that the predicted nonanalytic behavior is confined to just the first few percent drop in superfluid density, but we do not see why this should be so.

This work has been supported by NSF DMR grant 0203739. We acknowledge help from A. Weber, G. Hammerl, C.W. Schneider and J. Mannhart of the University of Augsburg, who made the sample D for this study.


\begin{thebibliography}{10}
\bibitem{Geshkenbein} V. B. Geshkenbein, L. B. Ioffe and A. I. Larkin, Phys. Rev. B, {\bf 55}, 3173 (1997)
\bibitem{Micnas} R. Micnas, S. Robaszkiewicz and A. Bussmann-Holder, Phys. Rev. B {\bf 66}, 104516 (2002)
\bibitem{Micnas1}R. Micnas, S. Robaszkiewicz and A. Bussmann-Holder, J. Supercond. {\bf 17}, p27-32, (2004)
\bibitem{Stajic} J. Stajic, A. Iyengar, K Levin, B. R. Boyce and T. R. Lemberger, Phys. Rev. B, {\bf 68}, 024520, (2003)

\bibitem{Kosztin} I. Kosztin, Q. Chen, Y.-J. Kao, and K. Levin, Phys. Rev. B, {\bf 61}, 11662 (2000)

\bibitem{Chen1} Q. Chen, I. Kosztin, B. Janko, and K. Levin Phys. Rev. Lett.{\bf 81}, 4708 (1998)

\bibitem{Chen2}Q. Chen and J. R. Schrieffer, Phys. Rev. B {\bf 66}, 014512 (2002)

\bibitem{Carrington} A. Carrington, I. J. Bonalde, R. Prozorov, R. W. Giannetta, A. M. Kini, J. Schlueter, H. H. Wang, U. Geiser and J. M. Williams, Phys. Rev. Lett., {\bf 83}, 4172 (1999)

\bibitem{Turneaure1} S.J. Turneaure, E.R. Ulm, and T.R. Lemberger, J. Appl. Phys. {\bf 79}, 4221 (1996)

\bibitem{Turneaure2} S.J. Turneaure, A.A. Pesetski, and T.R. Lemberger, J. Appl. Phys.  {\bf 83}, 4334 (1998).

\bibitem{Timusk} T. Timusk and B. Statt, Rep. Prog. Phys. {\bf 62}, 61 (1999).

\bibitem{Hirschfeld} P. J. Hirschfeld and N. Goldenfeld, Phys. Rev. B {\bf 48}, 4219 (1993).

\bibitem{Broun}D. M. Broun et al., cond-mat/0509223 (2006)
\end{thebibliography}
\end{document}